\documentstyle[sprocl]{article}
\input{psfig}
\def\be{\begin{equation}}
\def\ee{\end{equation}}
\def\bea{\begin{eqnarray}}
\def\eea{\end{eqnarray}}

\begin{document}
\title{Nucleon Instability in SUSY Models}
\author{ PRAN NATH}
\address{Department of Physics, Northeastern University,
Boston,\\ MA 02115, USA}
\author{ R. ARNOWITT }
\address{Center for Theoretical Physics, Department of Physics\\
Texas A \& M University, College Station, TX77843, USA}
\maketitle\abstracts{ A  review of the current status of nucleon 
stability in SUSY unified models is given. The review includes analysis
of nucleon stability in the minimal SU(5) supergravity model, and in 
extensions of the minimal model such as SO(10) models, and models
 including textures. Implications of the simultaneous  
 constraint of the experimental limit on the proton lifetime  and the
 constraint that dark matter not overclose  
 the universe are also discussed. }

 \noindent 
{\bf 1. ~~Introduction}\\

This review concerns the status of nucleon stability in supersymmetric 
	theories. First, we will give a general discussion of
	baryon number violation relevant to proton stability in SUSY
	theories. Next we will discuss in some detail the predictions
	of nucleon stability in the minimal supergravity model. Finally, we 
	will discuss proton stability in extensions of the minimal
	model and comment on the issue of proton stability in string models. 
	There are, of course, many sources of baryon number violation
	in unified theories.  First, grand unifed theories have 
	lepto-quark mediated proton decay and in SU(5) the proton decay via 
	lepto-quarks is given by\cite{gm} 
\begin{equation}
\tau(p\rightarrow e^+\pi^0) \approx (\frac{M_V}{3.5 \times 10^{14}  
GeV})^4 10^{31\pm 1} yr
\end{equation}
In SUSY SU(5) models one estimates  $M_V=1.1\times 10^{16}$ GeV, which gives
a p decay lifetime\cite{susy97} of $1\times10^{35\pm1}$. 
The current limit experimentally is\cite{pdg}  
\begin{equation}
\tau(p\rightarrow e^+\pi^0)>9\times  10^{32} yr, (90\% CL)
\end{equation}
and it is expected that Super Kamionkande (Super-K) and Icarus will be able to 
probe the $e^+\pi^0$ proton decay mode up to\cite{totsuka} $1\times 10^{34}$y. 
Thus the $e^+\pi^0$ mode in SUSY SU(5) may be on the edge of detection if
Super-K and Icarus reach their their maximum sensitivity. 

	 In supersymmetric theories there are  sources of proton
	 instability arising from terms in the superpotential. 
	 The main purpose of this talk is to review 
	  the status of proton stability in the presence of these 
	  additional sources. The outline of the rest of the paper is
	 as follows: In Sec.2 we discuss B and L 
	 violation in supersymmetric theories  from the superpotential
	 terms and show that generically even with R parity invariance  
	 there is p decay from dimension five operators in SUSY/string
	  unified models. In Sec. 3 we discuss p decay in 
	 the minimal SU(5) model. In sec. 4 we discuss p decay  in 
	 extensions of the minimal model. Conclusions are given in Sec. 5.\\

\noindent	 
{\bf 2. Sources of P Decay in SUSY/String Models}\\

	We turn now to a discussion of the sources B and L violating 
	interactions in the superpotential and their effect on proton 
	stability. It is well known that B and L
	violating  dim 4 operators
	 lead to fast proton decay. Thus, for example, the effective dimension
	 four low
	 energy $SU(3)\times SU(2)\times U(1)$  invariant interaction given by
	\begin{eqnarray}
	W=\lambda_u Qu^c H_2+\lambda_d Qd^c H_1+
	 \lambda_e Le^c H_1+\mu H_1H_2\nonumber\\
	+(\lambda_B' u^cd^cd^c + \lambda_L' Qd^cL+\lambda_L'' LLe^c)
	\end{eqnarray}
	The terms in the bracket lead to fast proton decay and  consistency 
	with experiment requires the constraint
	\begin{equation}
	\lambda_B'\lambda_L'< O(10^{-27})
	\end{equation}
	This type of decay is eliminated in the MSSM by the imposition of 
	R parity invariance [where $R=(-1)^{3B+L+2S}$].  
	It is likely that this discrete R symmetry is remnant of a global 
	continuos R symmetry. In that case there is the danger that the global
	symmetry may not be preserved by gravitational interactions. 
	For example, worm holes can generate baryon number violating
	 dimension 4 operators and catalize proton decay\cite{wormhole}. 
	A decay of this type is not suppressed by either a large exponential 
	suppression or
	 by a large heavy mass and is thus very rapid and dangerous. 
	 
		One way to protect against wormhole  induced fast proton decay 
		of the type discussed above is to elevate the relevant global
		symmetries which kill B and L violating dimension 4
	operators to gauge symmetries\cite{krauss,ir}. In this case if the local
		symmetry breaks down to a discrete symmetry, then the 
		left over discrete symmetry will be sufficient still to 
		protect againt  worm hole type induced proton decay. 
		There is, however, the problem that in unified theories 
		undesirable proton decay can arise as a consequence of
		spontaneous symmetry breaking even if the dangerous
		dimension 4 operators were  forbidden initially. Thus,
		for example, for SO(10) one can have a interaction of the
		type $(16)^4$ which gives terms of the following type in
		the superpotential 
		\begin{equation}
		\frac{1}{m_{Pl}}u^c d^c d^c\nu^c, \frac{1}{m_{Pl}}Q L d^c\nu^c
		\end{equation}
		If there is  a spontaneous generation of VEV for the 
		$\nu^c$ field, then terms of the type $u^cd^cd^c$ and
		terms of the type $QLd^c$ emerge which once again lead
		to a rapid  proton decay unless $<\nu^c>/M_{Pl}$ is 
		 O($10^{-13}$). Thus one must make certain that 
		 higher dimensional operators in the superpotential 
		 do not lead to dangerous p decay after
		 spontaneous symmetry breaking takes place or that 
		 spontaneous  symmetry breaking does not occur.

		Next we discuss p decay from dimension 5 operators
		(dimension 4 in the superpotential) which contain 
		B and L violating interactions. Using 
		$SU(3)$ $\times$ $ SU(2)$ $\times $ $U(1)$ invariance 
		one can write
		many B and L violating interactions. 
		Examples of such  
		terms in the superpotential involving matter fields are

		\begin{equation}
		QQQL,~~u^cu^cd^ce^c
		\end{equation}
		One can also have terms which involve the Higgs, e.g.,
			\begin{equation}
		         QQQH_1, Qu^ce^cH_1, etc.	
        		\end{equation}
	        The second set of terms can 
	        be eliminated if we  impose R parity invariance.
	        However, B and L violating terms of the first type arise
	        quite naturally in SUSY unified models via the exchange of
	        color Higgs triplet fields. In fact we now show that most
	        SUSY/string models will exhibit p instability via the 
	        Higgs triplet couplings. The  p decay arising
	        from the color Higgs exchange  is governed by 
	        the interactions\cite{testing}     	
\begin{equation}
\bar H_{1}J + \bar K H_1+ \bar H_{i} M_{ij} H_{j}
\end{equation}
where J and K are quadratic in the matter fields. 
It is then easily seen that the p decay from  dimension five operators is 
suppressed provided 
 \begin{equation}
(M^{-1})_{11}=0
\end{equation}
Now a condition of this type can be met in one of the following two ways:
 (i) discrete symmetries, and (ii) non-standard embeddings. However, 
 most SUSY/string models do not normally have discrete symmetries of the
 desired type which automatically satisfy Eq.(9),
 and only very specific SUSY/string models (the flipped 
 models\cite{flipped})
 satisfy (ii). Thus in general SUSY/string models do not have a natural 
 suppression of p decay via  dimension 5 operators. Thus suppression of
 p decay here  must be forced by making the Higgs triplets
 heavy.\\ 
 
 \noindent 
{\bf 3. ~~Proton Decay in Minimal SU(5) SUGRA Model}\\

	We begin with a discussion of the comparison of the minimal 
	SU(5) model predictions for the gauge coupling constants at the 
	electro-weak scale with the LEP data. It is well known that 
	the model with the MSSM spectrum extrapolated to high 
	energy gives a reasonably good fit for the coupling constants with 
	experiment\cite{deboer}.
	However, a more accurate analysis shows that the  theoretical 
	value of $\alpha_s$  predicted by SU(5) is
	about $2\sigma$ higher than experiment\cite{bagger1,das}. 
	There are a variety of 
	ways in which one can achieve a correction of size $2\sigma$.
	These include Planck scale corrections, and various other extensions
	of the minimal SU(5). We discuss here briefly the possibility of
	Planck scale corrections which are expected to be of size O(M/$M_{Pl}$)
	where M is size of the GUT scale. It is reasonable to expect that 
	such corrections are present due to the proximity of the 
	GUT scale to the Planck scale. 
	Effects of this size can arise  via corrections to
	 the gauge kinetic energy function\cite{das,hill,sarid}. 
	 For example, for the case of
	 SU(5) one can introduce a field dependence in the the gauge kinetic 
	 energy function scaled by the Planck mass so that 

	 \begin{equation}
	 f_{\alpha\beta} =(\delta_{\alpha\beta}+
	  \frac{c}{2M_P} d_{\alpha\beta\gamma}\Sigma^{\gamma}) 
	 \end{equation}
         Here  c parametrizes Planck physics and $\Sigma$ is the 24-plet
         of SU(5). The analysis shows that it is easy to generate a 
         2$\sigma$ correction to $\alpha_s$ with a   
         $c\sim 1$ to achieve full 
         agreement with experiment\cite{das}.
 	It is also possible to          
 	 understand a 2 $\sigma$ effect on $\alpha_s$ from corrections 
 	 arising from 
 	 extensions of the  minimal
	 SU(5) model such as, for example, in some modified versions of the
	  missing  doublet model\cite{tamvakis}.

Computation of the Higgsino mediated proton decay lifetime in supersymmetric 
theories involves both GUT physics as well as physics in the low energy
region via dressing loop diagrams which convert dimension five operators 
into  dimension six operators which can be used  for the computation
of proton decay amplitudes\cite{wein,acn}.
 Now the dressings involve 28 separate 
sparticle  masses and many trilinear couplings which mix left and right
squark fields. Thus is general no quantitative predictions of proton decay
can be made in the MSSM which has a large number of arbitrary parameters in it.
In the minimal SUGRA model\cite{can,applied} 
the number of arbitrary parameters is vastly 
reduced. Using the constraints of radiative breaking of the electro-weak 
symmetry one has only four arbitrary parameters and one sign in the SUSY
sector of the theory. These can be chosen to be 

\begin{equation}
m_0, m_{\frac{1}{2}}, A_0, tan\beta, sign(\mu)
\end{equation}
where $m_0$ is the universal scalar mass, $m_{\frac{1}{2}}$, is the universal
gaugino mass,  $A_0$ is the universal trilinear coupling, $tan\beta
=<H_2>/<H_1>$. Here $H_1$ gives mass to the down quark and $H_2$ gives mass
to the up quark, and $\mu$ is the Higgs mixing parameter. Because of the 
small number of parameters, the minimal supergravity model is very predictive.
In turns out that as a consequence of radiative breaking of the electro-weak
symmetry, over most of the parameter space of the minimal
supergravity model one finds that scaling laws hold and one has\cite{scaling}  
\begin{eqnarray}
m_{\tilde W_1}\sim \frac{1}{3} m_{\tilde g}~( \mu<0) ;
m_{\tilde W_1}\sim \frac{1}{4} m_{\tilde g}~(\mu>0)\nonumber\\
2 m_{\tilde Z_1}\sim m_{\tilde W_1}\sim m_{\tilde Z_2};
m_{\tilde Z_3}\sim m_{\tilde Z_4}\sim m_{\tilde W_2} >> 
m_{\tilde Z_1}\nonumber\\
m_H^0\sim m_A\sim m_{H^{\pm}}>>m_h
\end{eqnarray}

In the following we shall use the framework of supergravity to compute 
proton decay amplitudes. For concreteness we shall use the minimal SU(5) 
as the GUT group and  extensions to the non-minimal case will be 
discussed in the next section. 

	The interactions which govern proton decay in minimal SU(5) 
	are given by 

\begin{equation}
W_Y=-\frac{1}{8}f_{1ij}\epsilon_{uvwxy}H_1^uM_i^{vw}M_j^{xy}+
f_{2ij}\bar H_{2u}\bar M_{iv} M_j^{uv}
\end{equation}
After breakdown of the GUT symmetry and 
integration over the Higgs triplet field the effective dimension
five interaction below the GUT scale which governs p decay  
is given by\cite{acn}

\begin{eqnarray}
\it L^L_5= \frac{1}{M} \epsilon_{abc}(Pf_1^uV)_{ij}(f_2^d)_{kl}
( \tilde u_{Lbi}\tilde d_{Lcj}(\bar e^c_{Lk}(Vu_L)_{al}-
\nu^c_kd_{Lal})+...)+H.c.\nonumber\\
\it L^R_5= -\frac{1}{M} \epsilon_{abc}(V^{\dagger} f^u)_{ij}(PVf^d)_{kl}
(\bar e^c_{Ri}u_{Raj}\tilde u_{Rck}\tilde d_{Rbl}+...)+H.c.
\end{eqnarray} 
where   $L^L_5$ is the LLLL dimension five operator and   $L^R_5$
is the RRRR dimension five operator. The Yukawa couplings can be related 
to the quark masses at low energy by

 \begin{eqnarray}
m_i^u=f_i^u (sin2\theta_W/e)M_Z sin\beta\nonumber\\
m_i^d=f_i^d (sin2\theta_W/e)M_Z sin\beta
\end{eqnarray}
and $P_i$ are the inter generational phases given by 
\begin{equation}
P_i=(e^{i\gamma_i}), ~\sum_i \gamma_i=0; ~i=1,2,3
\end{equation}
There are many possible decay modes of the proton. The most dominant
of these are those which involve pseudo-scalar bosons and leptons. 
These  are  

\begin{eqnarray}
\bar\nu_iK^+,\bar\nu_i\pi^+ ; i=e,\mu,\tau\nonumber\\
e^+K^0,\mu^+K^0,e^+\pi^0,\mu^+\pi^0,
e^+\eta,\mu^+\eta
\end{eqnarray}
One can get an estimate of their relative strengths by the quark mass
factors and by the CKM factors that appear in their decay amplitudes. 
These are listed in Table 1. Additionally the decay amplitudes are
governed by relative contributions of the third  generation vs second
generation squark and slepton exchange in the loops. The relative contribution
of the third generation vs second generation exchange in the loops is 
governed by the ratios  $y^{tK}$, $y^{t\pi}$,..etc. These are also listed
in Table 1. 

\begin{center} \begin{tabular}{|c|c|c|c|}
\multicolumn{4}{c}{Table~1:~lepton + pseudoscalar decay modes of the proton
\cite{acn} } \\
\hline
SUSY Mode & quark factors  & CKM factors & 3rd generation enhancement\\
\hline
$\bar \nu_eK$ &$m_d m_c$  &$V_{11}^{\dagger}V_{21}V_{22} $
& $(1+y_1^{tK}) $\\
\hline
$\bar \nu_\mu K $ &$m_s m_c$  &$V_{21}^{\dagger}V_{21}V_{22} $
& $(1+y_2^{tK})$\\
\hline
$\bar \nu_\tau K $ &$m_b m_c$  &$V_{31}^{\dagger}V_{21}V_{22} $
& $(1+y_3^{tK}) $\\
\hline
$\bar \nu_e \pi,\bar \nu_e \eta $ &$m_d m_c$  &$V_{11}^{\dagger}V_{21}^2 $
& $(1+y_1^{t\pi}) $\\
\hline
$\bar \nu_\mu \pi,\bar \nu_\mu \eta$ &$m_s m_c$  &$V_{21}^{\dagger}V_{21}^2 $
& $(1+y_2^{t\pi}) $\\
\hline
$\bar \nu_\tau \pi,\bar \nu_\tau\eta $ &$m_b m_c$  &$V_{31}^{\dagger}V_{21}^2 $
& $(1+y_3^{t\pi}) $\\
\hline
$eK $ &$m_d m_u$  &$V_{11}^{\dagger}V_{12} $
& $(1+y_e^{tK}) $\\
\hline
$\mu K $ &$m_s m_u$  & & $(1-V_{12}V^{\dagger}_{21}-y_{\mu}^{tK}) $\\
\hline
$e\pi, e\eta $ & $m_d m_u$ &  
&  $(1-V_{11}V^{\dagger}_{11}-y_e^{t\pi}) $  \\
\hline
$\mu \pi,\mu \eta $ &$m_s m_u$  & $V_{11}^{\dagger}V_{21}^{\dagger}$
&     $(1+y_{\mu}^{t\pi}) $\\
\hline
\end{tabular} 
\end{center}
\vspace{0.5 cm}

Taking all these factors into account one can arrive at 
the following rough hierarchy of branching ratios. 
\begin{eqnarray}
BR(\bar \nu K)>BR(\bar \nu \pi)> BR(\it l K)> BR(\it l \pi)
\end{eqnarray}

  The most dominant decay mode of the proton normally is $\bar \nu K^+$.
This pattern can be modified in some situations because of the 
interference of the contributions from the third generation vs second 
generation making $\bar \nu \pi^+$. the most dominant decay mode. 
However, aside from this special situation which may 
correspond to fine tuning, the most dominant decay 
will be $\bar \nu K^+$. We discuss now the details of this decay mode
in minimal supergravity. The $p\rightarrow \bar\nu_iK^+$ decay width 
for the neutrino type $\nu_i$ is given by\cite{acn}  
\begin{equation}
\Gamma(p\rightarrow \bar\nu_iK^+)=(\frac {\beta_p}{M_{H_3}})^2|A|^2
|B_i|C
\end{equation}
Here $M_{H_3}$ is the Higgs triplet mass and $\beta_p$ is the matrix 
element between the proton and the vacuum state of the 3 quark operator so that
\begin{equation}
\beta_p U_L^{\gamma}=\epsilon_{abc}\epsilon_{\alpha \beta} <0|d_{aL}^{\alpha}
u_{bL}^{\beta}u_{cL}^{\gamma}|p>
\end{equation}
where $U^{\gamma}_L$ is the proton spinor.
The most reliable evaluation of $\beta_p$ comes from lattice gauge 
calculations and is given by\cite{gavela} 
\begin{equation}
\beta_p=(5.6\pm 0.5)\times  10^{-3} GeV^3
\end{equation}
The other factors that appear in Eq.(19) have the following meaning: A
contains the quark mass and CKM factors, $B_i$  are the functions
that describe the dressing loop diagrams, and C contains chiral Lagrangian
factors which convert a the Lagrangian involving quark fields to the
effective Lagraingian  involving mesons and baryons.
Individually these functions are given by   
\begin{equation}
A=\frac{\alpha_2^2}{2M_W^2}m_s m_c V_{21}^{\dagger} V_{21}A_L A_S
\end{equation}
where $V_{ij}$ are the CKM factors, and $A_L$ and $A_R$ are the long 
distance and the short distance 
 renormalization group suppression factors as one  evolves the
operators from the GUT scale down to the electro-weak 
scale.$B_i$ are given by 
\begin{equation}
B_i= \frac{1}{sin2\beta}\frac{m_i^d V_{i1}^{\dagger}}{m_sV_{21}^{\dagger}} 
[P_2 B_{2i}+\frac{m_tV_{31}V_{32}}{m_cV_{21}V_{22}} P_3B_{3i}]
\end{equation}
where the first term in the bracket is the contribution from the second
generation and the second term is the contribution from the third generation.
The functions $B_{ji}$ are the loop intergrals defined by

\begin{equation}
B_{ji}=F(\tilde u_i,\tilde d_j,\tilde W)+(\tilde d_j\rightarrow \tilde e_j)
\end{equation}
where 

\begin{eqnarray}
F(\tilde u_i,\tilde d_j,\tilde W)=[E cos\gamma_-sin\gamma_+\tilde f(\tilde
u_i,\tilde d_j, \tilde W_1)
+cos\gamma_+sin\gamma_-\tilde f(\tilde
u_i,\tilde d_j, \tilde W_1)]\nonumber\\
-\frac{1}{2} \frac{\delta_{i3}m_i^u sin2\delta_{ui}}{\sqrt 2 M_W sin\beta}
[E sin\gamma_-sin\gamma_+\tilde f(\tilde
u_{i1},\tilde d_j, \tilde W_1)
-cos\gamma_-cos\gamma_+\tilde f(\tilde
u_{i1},\tilde d_j, \tilde W_2)\nonumber\\
 - (\tilde u_{i1}\rightarrow \tilde u_{i2})]
\end{eqnarray}
and $\tilde f$ is given by 
\begin{equation}
\tilde f(\tilde u_i,\tilde d_j, \tilde W_k)=sin^2\delta_{ui}
\tilde f(\tilde u_{i1},\tilde d_j, \tilde W_k)
+cos^2\delta_{ui}
\tilde f(\tilde u_{i2},\tilde d_j, \tilde W_k) 
\end{equation}
with
\begin{equation}
f(a,b,c)=\frac{m_c}{m_b^2-m_c^2}[\frac{m_b^2}{m_a^2-m_b^2}ln(\frac{m_a^2}
{m_b^2})-(m_a\rightarrow m_c)]
\end{equation}
$ \gamma_{\pm}=\beta_+\pm\beta_- $,  
\begin{equation}
sin2\beta_{\pm}=\frac{(\mu\pm \tilde m_2)}{[4\nu_{\pm}^2
+(\mu\pm \tilde m_2)^2]^{1/2}}
\end{equation}
and 
\begin{equation}
\sqrt 2 \nu_{\pm}=M_W(sin\beta\pm cos\beta)
\end{equation}
\begin{equation}
sin2\delta_{u3}=-\frac{-2(A_t+\mu ctn\beta)m_t}{m_{\tilde t_1}^2-
m_{\tilde t_2}^2}
\end{equation}
\begin{eqnarray}
E=1~, sin2\beta>\mu\tilde m_2/M_W^2 \nonumber\\
~~~=-1,sin2\beta<\mu\tilde m_2/M_W^2  
\end{eqnarray}
Finally C is given by 
\begin{equation}
C=\frac{m_N}{32\pi f_{\pi}^2} [(1+\frac {m_N(D+F)}{m_B})
(1-\frac{m_K^2}{m_N^2})]^2
\end{equation}
where 
$f_{\pi}, D,F, ..$ etc are the chiral Lagrangian factors and 
 have the numerical values:
$ f_{\pi}=139$~MeV,D=0.76,F=0.48,$m_N$=938 ~MeV, ~$m_K$=495 ~MeV, 
and  ~$m_B$=1154.
 
The proton can also decay into vector bosons and these modes consist of 
\begin{eqnarray}
\bar\nu_iK^*,\bar\nu_i\rho,\bar\nu_i\omega ; i=e,\mu,\tau\nonumber\\
e K^*,\mu K^*,e\rho,\mu\rho,e\omega,\mu\omega
\end{eqnarray}
A similar analysis can be done for these modes\cite{yuan}.

We discuss now details of the analysis in minimal supergravity unification.
In the analysis one looks for the maximum lifetime of the proton as we span the 
parameter space of the minimal model within the naturalness constraints
which we take to be
\begin{equation}
m_0\leq 1 TeV, m_{\frac{1}{2}}\leq 1 Tev, tan\beta\leq 25
\end{equation}
In Fig. 1 we display the results of the maximum lifetime for the mode
$p\rightarrow \bar\nu K^+$ as a function
of the gluino mass. The analysis indicates that if 
Super-K and Icarus can reach the maximum sensitivity of $2\times 10^{32}$ y
for this mode\cite{totsuka,icarus} 
then most of the parameter space within the naturalness
limits indicated will be exhausted. The analysis including the dark matter
constraints was also carried out. In this analysis we impose the very
conservative constraint that the mass density of the relic neutralinos in 
the universe not exceed the critical relic  density needed to close
the universe. The results are also displayed in Fig 1. Here one finds that
this constraint limits the maximum gluino mass to lie below 500 GeV\cite{oak}. 
Most of this gluino mass range can be explored at the upgraded 
Tevatron  with an 
integrated luminosity of about\cite{kamon,amidei} 25$fb^{-1}$.\\ 

\begin{figure*}
\begin{center}
\hspace*{0.25in}
\psfig{figure=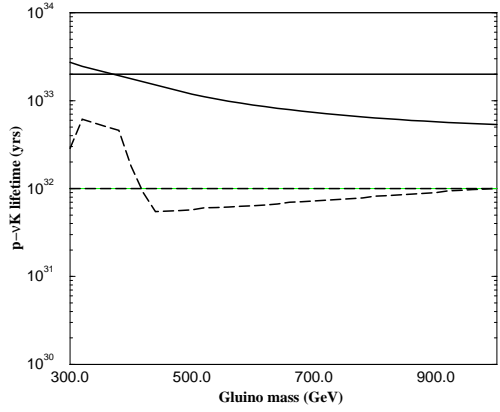,width=3.0in}
\end{center}
\caption[]{ 
Plot of the maximum $p\rightarrow \bar\nu K^+$ lifetime 
 under the constraint $m_0\leq 1$ TeV in minimal supergravity
model. The solid curve is for the case
when no relic density constraint is imposed.The dashed curve is with
the dark matter constraint $\Omega h^2<1$. 
The dashed horizontal line is the current
experimental limit and the solid horizontal line is the limit expected at  
Super-K and Icarus experiments.}
\label{fig:1}
\end{figure*}
\noindent 
{\bf 5. Nucleon Stability in Extensions of the Minimal Model}\\

	In Sec. 4 we discussed the situation regarding proton decay in
	the minimal SU(5) supergravity model. We discuss now the 
	situations in some extensions of the minimal model. There are various
	kinds of extensions of the minimal SU(5) model. These include 
	extension to larger groups, inclusion of textures, flipped 
	models, string models,  etc. We will consider here mostly
	the first two. In SO(10) one has a large  tan$\beta$, i.e.,
	  tan$\beta\sim 50$  to achieve  $b-t-\tau$ unification and 
	  compatibity with the observed  mass ratios for b, t and $\tau$.
	  However, a large tan$\beta$ tends to destabilize the 
	  proton\cite{urano}.
	  This can be seen from the fact that the current lower experimental 
	  limits for the proton lifetime for this decay mode  
	  	impose the following constraint on the effective 
	  	mass scale
	   $M_{PD}\equiv (M^{-1})_{11}$ 
         \begin{equation}
        M_{PD}> tan\beta (0.57\times 10^{16})  GeV 
        \end{equation}
        which for tan$\beta=50$ gives  $M_{PD}\sim 2.5 \times 10^{17}$ GeV.
       However, this  large scale tends to upset the unification of the
       gauge coupling constants\cite{urano,barr}. 
       The analysis
       of $\alpha_s$ vs $sin^2(\theta_W)$ is plotted in Fig.2 taken
       from ref.\cite{urano}. One 
       finds that for values of $M_{PD}$ of O($10^{17}$) GeV, the 
       disagreement of the theoretical value of $\alpha_s$ in minimal
       SO(10) with experiment is about $6\sigma$\cite{urano}. Thus one 
       needs large threshold corrections at the GUT scale to achieve 
       agreement with data\cite{lucas}.

\begin{figure*}
\begin{center}
\hspace*{0.25in}
\psfig{figure=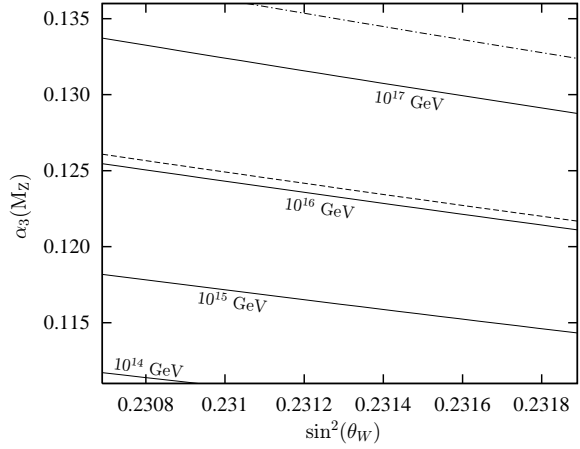,width=3.0in}
\end{center}
\caption[]{Contour plot of $M_{H_3}^{(0)}$ in $sin^2(\theta_W)$
 - $\alpha_3(M_Z)$ plane. The dashed line corresponds to the lower 
 bound on $M_{PD}$ in 
minimal SU(5) of $1.2\times 10^{16}$ GeV. The dot-dashed line 
corresponds to the lower limit on $M_{PD}$ in minimal SO(10) of 
$2.7\times 10^{17}$ GeV. Currently, the measurements are 
$sin^2(\theta_W)=0.2313\pm 0.0003$ and 
$\alpha_3(M_Z)=0.119\pm 0.0003$ (from ref.\cite{urano}). }
\label{fig:2}
\end{figure*}
 	Next we turn to another type of extension of the minimal SU(5) model
  	and it involves inclusion of textures to generate the correct 
  	quark-lepton mass hierarchies\cite{georgi}. 
  	One procedure to generate the
  	textures is to include Planck scale corrections in the
  	superpotential\cite{anderson,pn}, 
  	where the Planck scale corrections involve  expansions
  	in the ratio $\Sigma/M_{Pl}$, and  $\Sigma$  is the adjoint scalar 
  	field. After spontaneous breaking of the GUT group $\Sigma$ 
  	develops a VeV and one gets a hierarchy in the ratio $M/M_{Pl}$
  	which allows one to generate the textures in the Higgs doublet sector
  	defined by 
  	\begin{equation}
	W_{d}=H_1 l A^Ee^c+H_1d^c A^Dq+ H_2u^c A^Uq
	\end{equation}
	where $A^E$, $A^D$ and $A^U$ are the textures matrices.
  	Once the textures in the Higgs doublet sector are fixed one can
  	compute the textures in the Higgs triplet sector defined  by
  	\begin{eqnarray}	
	W_t=H_1 l B^E q + H_2 u^c B^U e^c
	+\epsilon_{abc} (H_1 d^c_b B^D u^c_c + H_2^a u^c_b C^U d_c)
	\end{eqnarray}	
  	where $B^E$, $B^U$, $B^D$, and $C^U$ are the triplet textures.

  	For the case when
  	one adds the most general Planck scale interaction  one finds that
  	fixing the textures in the Higgs doublet sector does not uniquely
  	fix the textures in the Higgs triplet sector\cite{pn}. 
  	One needs a dynamical
  	principle to do so. One possibility suggested is to extend 
  	supergravity models to include not only the usual visible and hidden
  	sectors, but also an exotic sector\cite{pn}. 
  	The exotic sector contains new
  	fields which transform non-trivially under the GUT group and
  	couple to the fields in the hidden sector and the adjoint scalars
  	in the visible sector. After spontaneous supersymmetry breaking the
  	exotic fields develop Planck scale masses because of their couplings
  	with the hidden sector fields and integration over the exotic fields
  	 leads to the desired Planck scale corrections. For the choice  of a 
  	 minimal set of exotic fields one finds that the Planck scale
  	 corrections are uniquely determined and one finds that correspondingly
  	 the textures in the Higgs triplet sector are uniquely determined.
  	  
  		We give now some specifics. 
 		The simplest examples of textures are the 
 		Georgi-Jarlskog matrices given by
\begin{equation}
A^E=\left(\matrix{0&F&0\cr
                  F&-3E&0\cr
                  0&0&D\cr}\right),
A^D=\left(\matrix{0&Fe^{i\phi}&0\cr
                  Fe^{-i\phi}&E&0\cr
                  0&0&D\cr}\right),                 
A^U=\left(\matrix{0&C&0\cr
                  C&0&B\cr
                  0&B&A\cr}\right)
\end{equation}	
If one uses the most general Planck scale interaction then fixing the
texture in the Higgs doublet sector to be the Georgi-Jarlskog does not fix 
the textures uniquely in the Higgs triplet sector.
However, for the
case when one uses the exotic sector hypothesis with the minimal set of
exotic fields one finds that the textures in the Higgs sector are uniquely
determined and are given by\cite{pn}

\begin{equation}
B^E=\left(\matrix{0&aF &0\cr
  a^*F &{16\over 3}E &0\cr
                  0&0&{2\over 3}D\cr}\right), 
B^D=\left(\matrix{0&-{8\over 27}F &0\cr
       {-8\over 27}F &-{4\over 3}E &0\cr
                  0&0&-{2\over 3}D\cr}\right)
\end{equation}

\begin{equation}
B^U=\left(\matrix{0&{4\over 9}C &0\cr
                  {4\over 9}C &0&-{2\over 3}B\cr
                  0&-{2\over 3}B&A\cr}\right)
\end{equation}
where a=$(-{19\over 27}+e^{i\phi})$  and $C^U=B^U$. 
Estimates including the textures show modest (i.e. O(1)) modifications 
in the p decay lifetimes. Further, the various p decay branching ratios
are affected differentially and thus measurement  of the branching ratios
will shed light on the textures and on the nature of physics at the GUT scale. 

	Finally, we discuss briefly proton stability in   string
	models. It is difficult to make general remarks here as 
	there are a huge variety of string models and further the 
	mechanism of supersymmetry breaking in string theory is not fully
	understood and thus the nature of soft SUSY breaking parameters
	is not known. However, there are various types of parametrizations
	that have been used in the literature. One such parametrization is
	that of no-scale models where one assumes that at the GUT scale
	one has $m_0=0=A_0$ and radiative breaking of the electro-weak
	symmetry is driven by $m_{\frac{1}{2}}$ and the top mass $m_t$. 
	In this case it is shown that the minimal SU(5) would lead to 
	a proton decay via dimension five operators which would be  
	in violation of the existing experimental bounds for the $\bar \nu K^+$
	mode\cite{noscale}. 
	This difficulty is  eliminated in the flipped no-scale 
	models\cite{flipped} where the $\bar\nu K^+$ mode is 
	highly suppressed.   	 
	Of course, as mentioned above there is a huge array
	 of string models and each model must be individually examined to
	test p stability. Thus an analysis of p stability in three generation
	Calabi-Yau models can be found in ref\cite{calabi}. Similarly an
	analysis of p stability  in superstring derived
	standard like models is given in ref. \cite{farragi}.\\
{\bf 7.Conclusion}\\

In this review we have given a brief summary of the current status of
proton instability in SUSY, SUGRA and string unified models. One of the 
important 
conclusions that emerges is that if the Super-K and Icarus experiments 
can reach the expected sensitivity of  $2\times 10^{34}$ y for the 
$\bar\nu K^+$ mode, then one can probe a majority of the parameter space 
of the minimal SUGRA model within the naturalness constraint of 
$m_0\leq 1$ TeV,  $m_{\tilde g}\leq 1$ TeV, and $tan\beta\leq 20$. 
If one includes the additional
constraint that the neutralino relic density not exceed the critical 
matter density needed to close the universe then the gluino mass must lie
below 500 GeV  to satisfy the 
current lower limit on the $\bar \nu K^+$  mode. These results will be 
tested in the near future in p decay experiments as well by 
 experiments at the Tevatron, LEP2 and the LHC. \\

\noindent 
{\bf Acknowledgements}:\\

This research was supported in part by NSF grant numbers 
PHY-9602074  and  PHY-9722090.\\

\end{document}